\documentclass[a4paper]{article}

\usepackage{INTERSPEECH2021}
\usepackage[linesnumbered, ruled]{algorithm2e}
\usepackage{comment}
\usepackage{subfigure}
\usepackage{multirow}
\usepackage{multicol}
\usepackage{amsmath}
\usepackage{tabularx} 
\usepackage[utf8]{inputenc} 
\usepackage[T1]{fontenc}    
\usepackage{hyperref}       
\usepackage{url}            
\usepackage{booktabs}       
\usepackage{amsfonts}       
\usepackage{nicefrac}       
\usepackage{microtype}      
\usepackage{xcolor}         
\usepackage{graphicx}
\usepackage{subcaption} 
\usepackage{amsfonts}       
\usepackage{amsmath}        
\usepackage{cite}

\title{Explore the Reinforcement Learning for the LLM based ASR and TTS system}
\name{Changfeng Gao, Yabin Li, Keyu An, Zhifu Gao, Zhihao Du, Han Zhao, Xiangang Li}
\address{Speech Team, Tongyi Lab, Alibaba Group}
\email{\{gaochangfeng.gcf\}@alibaba-inc.com}
\begin{document}

\maketitle
\begin{abstract}

In recent years, large language models (LLMs) have played an important role in automatic speech recognition (ASR) and text-to-speech (TTS) systems. While reinforcement learning (RL) has significantly enhanced LLM performance in text-based tasks, its application to ASR and TTS remains underexplored due to the complexity of training audio-based models.
In this study, we propose a lightweight RL framework tailored for audio-based LLMs that can process audio inputs and generate audio outputs. Based on this framework, we evaluate the effectiveness of reinforcement learning on both ASR and TTS tasks.
For the ASR task, we experiment with different rule-based reward functions within the Group Relative Policy Optimization (GRPO) framework and investigate the impact of RL data construction.
For the TTS task, we compare GRPO with Differentiable Reward Optimization (DiffRO) and further combine the two approaches to achieve improved performance.
Our experiments demonstrate that RL can significantly enhance the performance of both ASR and TTS systems, even with limited training data and a small number of optimization steps. 

\end{abstract}


\section{Introduction}

The development of large language models (LLMs) has significantly influenced the speech domain, including both automatic speech recognition (ASR) and text-to-speech (TTS). Most recently published ASR and TTS systems are now based on LLMs, benefiting from scaling in data and model size. These systems demonstrate substantial advantages over traditional small neural network models.

In LLM training, reinforcement learning (RL) is a critical step that aligns models with human preferences and enhances their reasoning capabilities \cite{ziegler2020finetuninglanguagemodelshuman}. During RL, the model generates responses to a given input, and these responses are evaluated using either a learned reward model or handcrafted rules. The policy is then optimized to maximize the expected reward.
Compared to supervised fine-tuning (SFT), RL directly shapes the model’s outputs to satisfy predefined rules or human preferences, rather than merely maximizing the posterior probability of the training labels.

Several studies have applied RL to audio-based LLM systems, including ASR and TTS. Among recent audio LLM-based systems, Step-Audio2\cite{wu2025stepaudio2technicalreport} investigates the effectiveness of RL by employing both PPO\cite{schulman_proximal_2017} and GRPO\cite{shao2024deepseekmathpushinglimitsmathematical} to improve performance. For ASR, Seed-ASR\cite{bai2024seedasrunderstandingdiversespeech} demonstrates that minimum word error rate (MWER)\cite{prabhavalkar2017minimumworderrorrate} training can slightly reduce WER and improve the F1 score on hard cases. Similarly, Seed-TTS\cite{anastassiou_seed-tts_2024} shows that REINFORCE, PPO, and DPO\cite{rafailov_direct_2023} can enhance TTS performance in terms of speaker similarity and WER. In CosyVoice3\cite{du2025cosyvoice3inthewildspeech}, differentiable reward optimization (DiffRO)\cite{gao2025differentiablerewardoptimizationllm} is used to align the discrete tokens predicted by the TTS system with the preferences of a multi-task reward model, resulting in significant improvements in WER—indicating that RL is an essential component in TTS training. Moreover, some works have shown that RL can also improve a model’s ability to express emotion and follow instructions.

However, compared to text and video modalities, RL research for audio LLMs remains limited. Most existing studies only report which RL methods were used, often without detailed effectiveness analysis or ablation studies on key aspects such as reward rule design, reward model training, or RL data construction. Furthermore, compared to video LLMs, audio LLMs are more flexible and complex: they can process both discrete acoustic tokens and continuous embeddings as input, and generate either acoustic tokens or text tokens as output.

In this paper, we conduct a detailed investigation into the effectiveness of RL in LLM-based ASR and TTS systems. We first design a lightweight RL framework for audio LLMs that can take audio embeddings as input and efficiently generate text tokens or synthesize waveforms as output. For ASR, we compare traditional MWER with GRPO under different reward functions and RL data construction strategies. For TTS, we compare GRPO with DiffRO and further combine the two methods to achieve additional performance gains.
Experiments show that each RL method contributes to performance improvement, even with a small amount of training data and few optimization steps. Crucially, reward design and RL data construction significantly influence the model’s output preferences, ultimately affecting user experience.


\section{Related Work}

\subsection{Group Relative Policy Optimization (GRPO)}

GRPO is a lightweight and effective policy-based method that has realized great success for the LLM post-training. Unlike other RL algorithms like PPO, GRPO eliminates a group of response $\{o_i\}_{i=1}^{G}$ with some rule-based value function to get the reward $\{R_i\}_{i=1}^G$ and then normalize the group-level rewards to get the advantage $\hat{A}_{i,t}$:

\begin{equation}
\hat{A}_{i,t}=\frac{R_i-\text{mean}\!\left(\{R_j\}_{j=1}^G\right)}{\text{std}\!\left(\{R_j\}_{j=1}^G\right)}
\end{equation}

The policy will be optimized with a clipped objective and a directly imposed KL penalty term.

\begin{equation}
\begin{split}
L_{\mathrm{GRPO}}(\theta) = 
\frac{1}{G}\sum_{i=1}^G \frac{1}{|o_i|}\sum_{t=1}^{|o_i|} \left[  \hat{r}(\theta) - \beta\,D_{\mathrm{KL}}\!\bigl(\pi_\theta\|\pi_{\mathrm{ref}}\bigr) \right]
\end{split}
\end{equation}
where
\begin{equation}
\hat{r}(\theta)= \min\!\bigl(r_{i,t}(\theta)\,\hat{A}_{i,t},\;
\operatorname{clip}\bigl(r_{i,t}(\theta),1-\varepsilon,1+\varepsilon\bigr)\,\hat{A}_{i,t}\bigr)
\end{equation}
and
\begin{equation}
r_{i,t}(\theta)=\frac{\pi_{\theta}\bigl(o_{i,t}\mid q,\,o_{i,<t}\bigr)}
{\pi_{\theta_{\mathrm{old}}}\bigl(o_{i,t}\mid q,\,o_{i,<t}\bigr)}.
\end{equation}

\subsection{Differentiable Reward Optimization (DiffRO)}

DiffRO is a recent published RL algorithm whch is desgined for the LLM based TTS system. Compare the other RL method, DiffRO can directly predict the reward from the generated token rather than the synthetic audio, and directly optimize the TTS system with the gradient back-propagated from the reward model. 
It first use the GumbelSoftmax to sample the output token $\tilde{o}_t$ predicted from the LLM, and then feed the sampled result into a token-based ASR system, then the ASR system calculate the posterior probability of the input text $y_{1:N}$ and then use this probability $ P_{ASR}$ as reward to optimized the TTS model by  back-propagation to maximin the posterior probability.

\begin{equation}
\tilde{o}_t = \text{GumbelSoftmax}_{\text{hard}} P_{\pi_{\theta}}(o_t | o_{1:t-1}; y_{1:N})   \label{eq:gumbel}
\end{equation} 
\begin{equation}
    R_{ASR}(Y) = \log P_{ASR}(y_{1:N} | \tilde{o}_{1:T})   \label{eq:diffro}
\end{equation}
Besides the ASR reward, DiffRO can also use other downstream tasks to guide the TTS model training like speaker emotion and speech quality.

\begin{figure*}[!t]
    \centering
        \centering
        \includegraphics[width=0.7\linewidth]{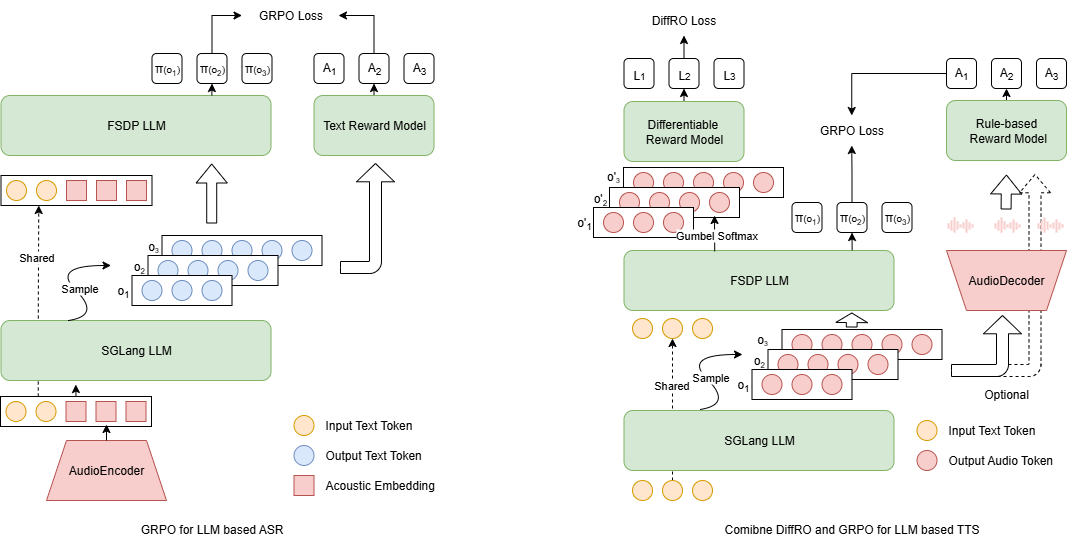} 
        \label{fig:rl_left}
    \caption{Training framework and time consumption analysis for the reinforcement learning.}
    \label{fig:combined}
\end{figure*}

\subsection{Training Frameworks for RL}

Compared to supervised fine-tuning (SFT), reinforcement learning (RL) training is significantly more complex, as it requires managing computational resources across multiple components—such as the Actor, Reward model, Critic, and Reference model—and establishing efficient communication for training and synchronization.
TRL\cite{vonwerra2022trl} provides a simple implementation of RL but is less efficient; it relies directly on PyTorch for both training and inference and uses manual resource management.

In contrast, more advanced industrial-scale frameworks such as VeRL\cite{sheng2024hybridflow} and OpenRLHF \cite{hu2024openrlhf} leverage Ray’s flexible distributed computing primitives to streamline distributed operations and deployment. These frameworks further integrate high-efficiency rollout engines like VLLM and SGLang for response generation, and employ Deepspeed/FSDP with ZeRO-based strategies for policy optimization.
With these system-level optimizations, they can significantly accelerate RL training speed and reduce GPU memory consumption.

\section{RL Framework for the Audio LLM}

Unlike vanilla LLMs, audio LLMs typically include an audio encoder (for ASR) or decoder (for TTS). These frontend or backend modules add further complexity to the training pipeline and inter-process communication. To simplify the management of computational resources, we design a RL training framework that alternately allocates GPU resources among different components. The overall framework is illustrated in Figure~\ref{fig:combined}.

For ASR training, a vanilla PyTorch-based audio encoder first occupies the GPU. It processes all input audio samples in a single batch, extracting audio embeddings in parallel. Once completed, it releases the GPU resources and transfers the embeddings to main storage.
Next, an SGLang-based LLM rollout engine takes control of the GPU to generate multiple groups of hypotheses based on the audio embeddings and instruction text tokens. Each hypothesis is assigned a reward according to predefined rules for advantage computation.
Finally, an FSDP-based LLM policy model uses the audio embeddings and generated hypotheses to compute output probabilities and performs policy optimization. After each update, the updated policy is synchronized back to the rollout engine to ensure on-policy training.

For TTS training, the SGLang-based engine directly generates acoustic token sequences, which are then fed into a PyTorch-based flow-matching model and vocoder to produce Mel-spectrograms and waveforms. Subsequently, GPU resources are allocated to a PyTorch-based reward model (when used) to compute reward scores. The FSDP-based policy model then leverages these reward scores and the generated token sequences to perform reinforcement learning using either GRPO or DiffRO.

\section{GRPO for ASR}

\subsection{Reward Rules Design}

Although the LLM based ASR can realize a good results in most case, but it can could change the keyword or fall in hallucinations which serious impact on the user experience. To address these hard case samples, we design some rule based value functions $\{R^k(y_i^*, y_i)\}_{k=1}^K$ for the GRPO based RL to improve the ASR performance more than reduce word error rate (WER).

\begin{enumerate}
    \item $R^1$: ASR Accuracy. To improve the ASR performance, we use the $1 - \text{WER}(y^*, y)$ as the basic value function.
    \item $R^2$: Hallucination Detection. Hallucination is a common problem of the LLM-based ASR system and will be more serious on noisy data; it could generate repeated, inexistent, and translated results. We detach these hallucination by some rules and it occurred, the final reward will be set to -1.
    \item $R^3$: Keyword Accuracy and Recall. As the keyword has more impact on the user experience, we use the average between the recall and precision rate for the keyword as reward.
\end{enumerate}

\subsection{Training Data Collection} \label{sec:data_asr}

Addressing practical issues in application scenarios, we build a small but high quality training dataset for the RL with the approach. We first collect the hard and hallucination-related samples $D^1$ by compare transcription between different ASR systems, choosing the audios whose outputs are different or contain long repetitions. We also select audio segments longer than 20 seconds which are limited in the training dataset as $D^2$. For $D^3$, we choose the speech contains name, brand and other specialized vocabulary as the keywords samples. Finally, we also construct a random selection set $D^0$ as a control.

\section{Combine DiffRO and GRPO for TTS}

Previous works has shown that the DiffRO can significant improve the pronunciation accuracy of the TTS system and the speech attribute control with different downstream task. However, it relay on a differentiable neutral network based reward model to compute the reward and optimize the TTS model. But for some subjective experience, it is difficult to train the reward model like the sound expressiveness sound, rhythm richness and audio duration. 
GRPO can measure a part of the subjective experience by some rules which means that the GRPO and the DiffRO can combine with each other for a better performance.

\subsection{Rewards Desgin for GRPO}

Similarly to ASR, we also define some rules $\{R^k\}_{k=1}^K$ to compute the rewards of the responses. Aseptically, some  reward value can be computed according to the acoustic token, while some rules can be only applied on the waveform.

\begin{enumerate}
\item $R^1$: Recognition Accuracy of the ASR system. We can directly use the ASR recognition accuracy as the reward to prevent the error pronunciation. The accuracy can be calculated by a regular ASR from the synthesized audio or a token based ASR model like DiffRO.
\item $R^2$: Audio Duration. Some works has find that during RL, the TTS system tend to slow the speech speed for better ASR result. To prevent this problem, we use the difference from median audio length in the response group as reward:
\begin{equation}
    R_i^2 = - \text{abs}\left(\frac{|o_i| - T_{m}}{T_{m}} \right)
\end{equation}
\item $R^3$: Token and Pitch Diversity. We encourage the LLM generate more diversity acoustic token to increase expressiveness. The token diversity is calculated by the edit distance between the respones in one group, and the pitch diversity are measure by the standard deviation of the normalized $F_0$
\begin{equation}
    R_i^3 = \frac{1}{G} \sum_{j=1}^{G}\frac{\text{dist}(o_i, o_j)}{|o_i|}  + \text{std}(F_0)
\end{equation}

\end{enumerate}

\subsection{Combine GRPO and DiffRO by Sample Filter}

After get the GRPO reward, we can distinguish the response into positive ($\hat{A}_i > 0$) and negative ($\hat{A}_i \le 0$) samples according to the advantages. 
With RL, the posterior probability of the positive responses should increase while the negative's should be opposite. This optimization could be difficult and unstable. To address this problem, we also compute the DiffRO loss on the positive sample like eq \ref{eq:diffro}. This means that the combined RL method can not only increase the probability of the positive response but also tell the model how to make them better, so the training process will be more effective and stable.

\

\section{Experiments}

\subsection{Training Efficiency Analysis}
\begin{figure}
    \centering
    \includegraphics[width=0.85\linewidth]{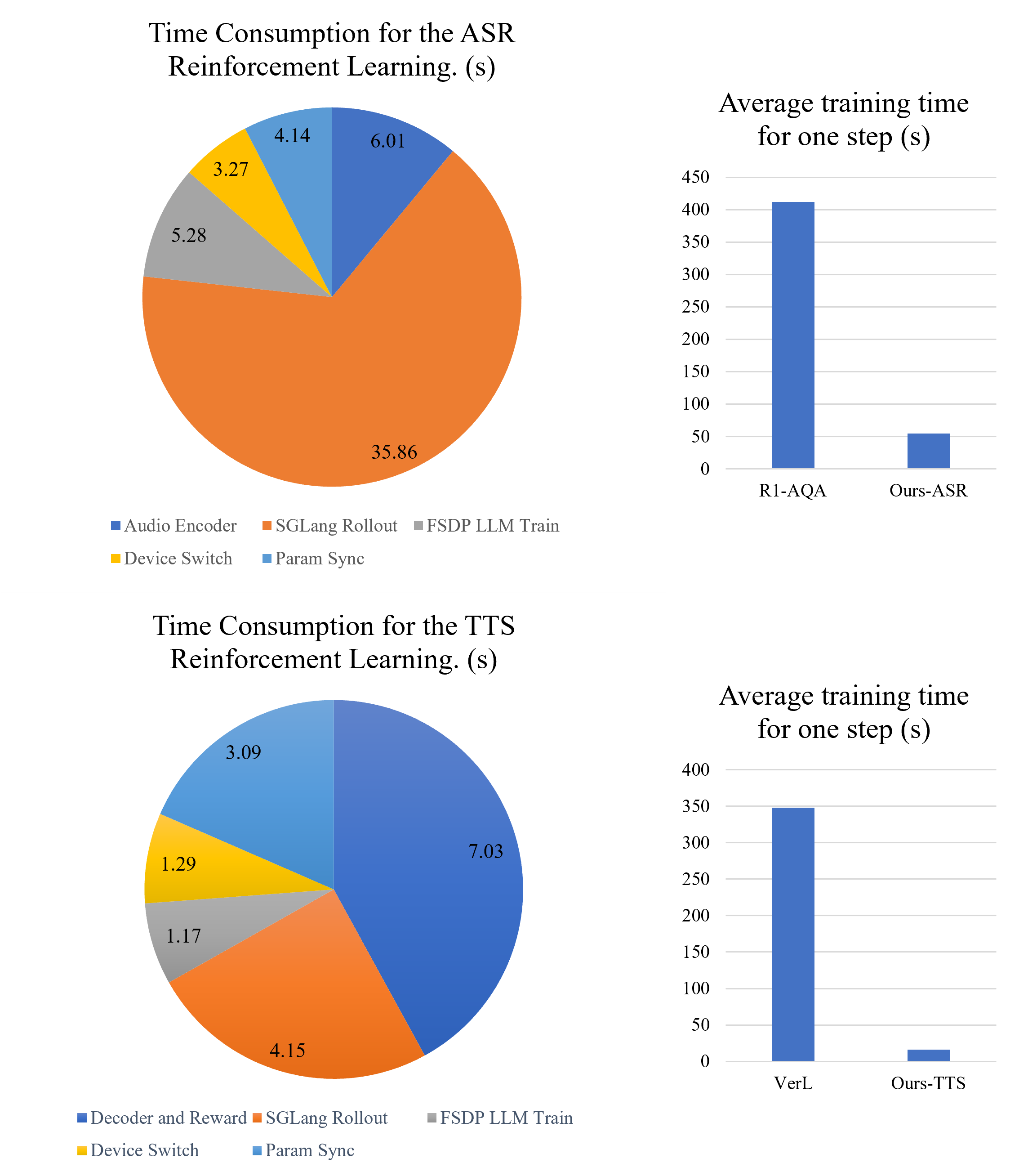}
    \caption{Training Effective Analysis.}
    \label{fig:time}
\end{figure}

We analyze the training efficiency of training frame works on 8 A100 GPUs and show the result in Figure~\ref{fig:time}. For the ASR, the experiments are based on the FunAudio-ASR\cite{an2025funaudioasrtechnicalreport}, which contains a 0.7B encoder and a Qwen2.5-7B based LLM. The total duration of the input audio is about an hour, and one training step consumes about 54.6 seconds. So the real-time factor (RTF) is about 0.015. And for the TTS, the experiments are based on the CosyVoice2-0.5B\cite{du2024cosyvoice2scalablestreaming}. We can see that the FM decoder will consume the most time and total duration for one training step is 16.73s with 128 batchsize. And for both ASR and TTS, the time consumption of the device switch and the parameters synchronization only take a small part. We further compare the RL some opensource RL frame work for Qwen2-Audio (R1-AQA)\cite{r1aqa} and Cosyvoice2 (VeRL based)\footnote{https://github.com/nvidia-china-sae/mair-hub/tree/main/rl-tutorial/cosyvoice\_llm}, and find our frame work has a significant advantage on the training speed. As the training is based on same LLM backbone, we believe that our alternate GPU utilization framework is a high effective training method for audio LLM.

\subsection{Experiments on ASR}

\subsubsection{Experiments Setup}

The ASR RL experiments are based on FunAudio-ASR. For training data construction, we collect the training set as described in Section~\ref{sec:data_asr}, with each subset  $\{D_i\}_{i=0}^3$ contributing 20k utterances.
For the training configuration, the batch size is set to 32 and the group sample size to 12. The LLM policy is trained online with a learning rate of 0.00001, and the KL divergence coefficient between the policy and the reference model is set to 0.1. With this setup, the training process can be completed within one day.
For evaluation, we use two in-house industrial test sets containing short (less than 10s) and long (longer than 20s) audio samples. To better analyze the hallucination problem, we report not only the overall WER but also insertion (Ins) and deletion (Del) error rates.

\subsubsection{Experiments Reults}

We present the ASR experimental results in Table~\ref{tab:asr_result}. From the table, we observe that RL achieves a relative improvement of 5.3\% in WER on both short and long speech segments. Moreover, the design of training data and reward functions plays a crucial role in RL effectiveness, with distinct effects observed on the short and long evaluation sets.
For short audios, all data construction strategies and reward designs improve ASR performance, and combining them yields the best overall result.
However, for long audios, we observe that even the baseline model exhibits significantly higher insertion (Ins) errors compared to deletion (Del) errors, indicating a tendency toward hallucination—generating words not present in the input. Simply including more long audio samples in training or adding hallucination detection mechanisms does not effectively address this issue. In contrast, using hard training samples $D^1$ proves beneficial, suggesting that RL training data should be carefully constructed based on the failure patterns of the base model.
Furthermore, keyword-related data and reward design are critical for mitigating hallucinations, as they impose stronger penalties than standard WER-based rewards when the model generates non-existent or irrelevant words.



\begin{table}[!t]
    \centering
    \begin{tabular}{ll cccccc}
    \hline
    \multirow{2}{*}{$R$} &  \multirow{2}{*}{$D$} & \multicolumn{3}{c}{Short}& \multicolumn{3}{c}{Long}\\
       & & WER & Ins & Del & WER & Ins & Del  \\
    \hline
     - & - &  10.25 & 2.51 & 2.25 & 6.35 & 2.72 & 0.87 \\
    $R^1$ & $D^{0}$ & 10.17 & 2.61 & 2.04 & 6.63& 3.05 & 0.87 \\
    $R^{1,2}$ & $D^{0}$ & 10.14 & 2.63 & 1.96 & 6.54 & 3.01 & 0.84 \\
    $R^{1,2}$ & $D^{0,2}$ & 10.08 & 2.60 & 2.08 & 6.35 & 2.86 & 0.69 \\
    $R^{1,2}$ & $D^{1,2}$ & 10.01 & 2.67 & 1.87 & 6.25 & 2.69 & 0.82 \\
    $R^{all}$ & $D^{all}$ & 9.71 & 2.61 & 1.68 & 6.03 & 0.86 & 0.75 \\

    \hline
    \end{tabular}
    \caption{Comparison for ASR models with different reward rules and data.}
    \label{tab:asr_result}
\end{table}

\subsection{Experiments on TTS}

\begin{figure}
    \centering
    \includegraphics[width=0.85\linewidth]{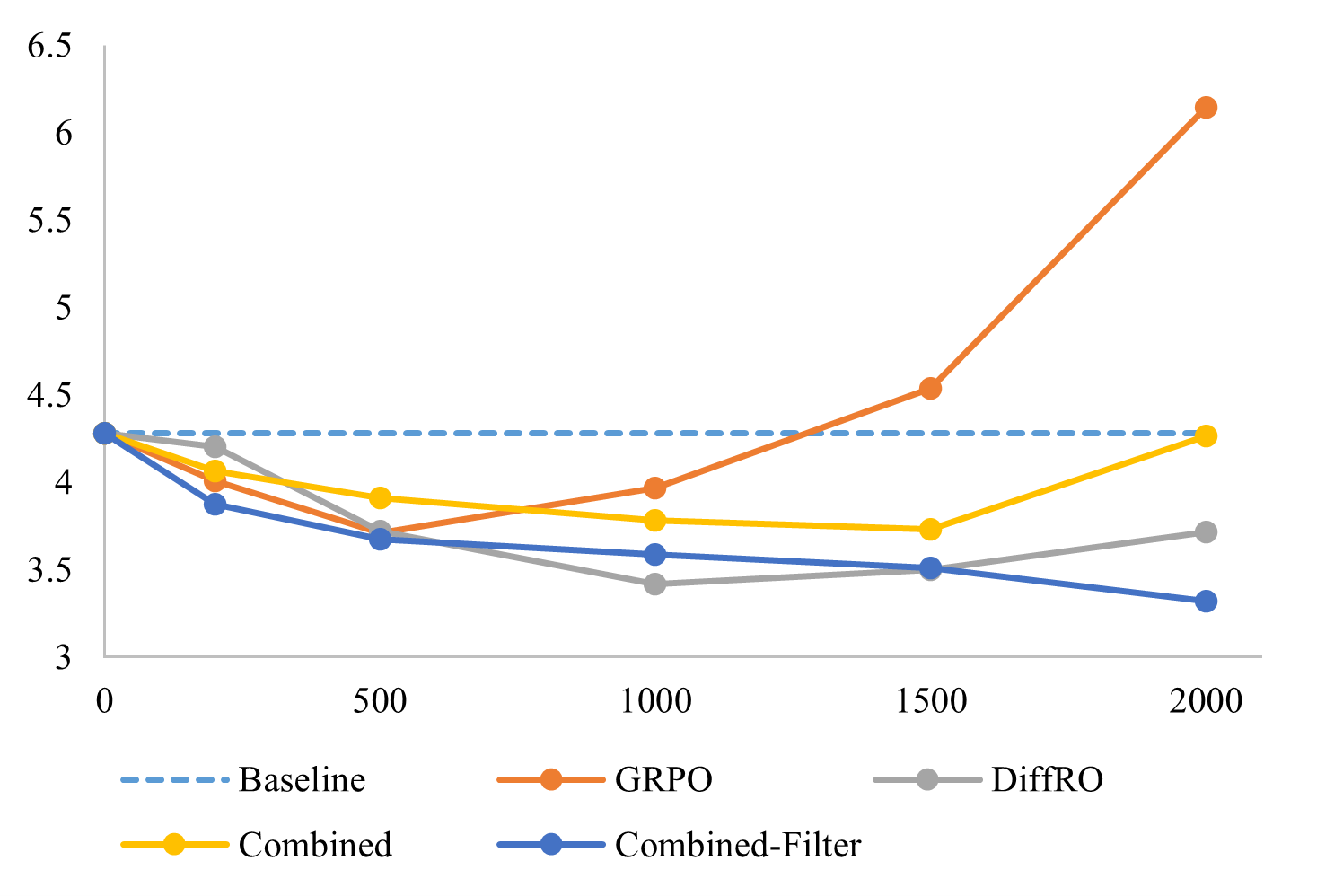}
    \caption{Training Stability Analysis for different RL methods.}
    \label{fig:rl_step}
\end{figure}

\begin{table}[!t]
    \centering
    \begin{tabular}{ll cc cc}
    \hline
    \multirow{2}{*}{Method} &  \multirow{2}{*}{Reward} & \multicolumn{2}{c}{zh}& \multicolumn{2}{c}{en} \\
       & & WER & SS & WER & SS \\
    \hline
    - & - & 4.280 &	77.64 & 6.074 & 70.87 \\
    DiffRO & $R^1$ & 3.418 & 77.00 & 5.690 & 70.01 \\
    GRPO & $R^1$ & 3.710 & 77.26 & 5.974 & 70.26 \\
    Combined & $R^1$ & 3.782 & 77.04 & 6.239 & 70.13 \\
    + Filter & $R^{1}$ & 3.381 & 76.73 & 5.401 & 69.97 \\
    + Filter & $R^{1,2}$ & 3.330 & 77.37 & 5.279 & 70.41 \\
    + Filter & $R^{1,2,3}$ & 3.414 & 77.11 & 5.579 & 70.05 \\

    \hline
    \end{tabular}
    \caption{Accuracy comparison for Audios synthesized from Different TTS system.}
    \label{tab:tts_result}
\end{table}

\subsubsection{Experiments Setup}
For the TTS experiments, we use the open-source CosyVoice2-0.5B as the base model and construct the training set using text data from CommonVoice zh and en.
We compare DiffRO and GRPO under various reward settings and explore different strategies for combining them. The reward model follows the setup in~\cite{gao2025differentiablerewardoptimizationllm}, and the KL divergence coefficient is set to 0.1.
All experiments are conducted with a batch size of 16. When GRPO is used, the group sample size is set to 8 and the sampling temperature is 1.0.
Evaluation is performed on the CV3-Eval\cite{du2025cosyvoice3inthewildspeech} test set.

\subsubsection{Experiments Reults}

We show the experiments results in Tabel \ref{tab:tts_result} and can find that most RL method can improvement the WER but slightly harm the speaker similarity (SS). This is reasonable as the CosyVoice2 only use the semantic information in the LLM and the RL reward also focus on the semantic and rhythm.
And when only consider the WER, the DiffRO can be better than the GRPO, but the SS reduction is also larger. And if we directly combine them together, the result can become even worse. However, sample filter based combination can fix this problem and realize a much better result. Because we find that when sample a response group with high temperature, some response could contains many repeated token can not predict the stop token. Directly compute the differentiable reward loss on these bad sequence can be harmful. So filter these bad case can make GRPO and DiffRO more compatible. 
Based on the sample filter  combination, we explore more reward function in the GRPO, and find that the audio duration reward can further improve the performance, and it can prevent the speech speed come slower.  The diversity can not improve the objective WER and SS result, but for the subjective evaluation, it achieve the best result than others.

Besides the performance, we also compare the training stability between different RL method and show the results in Figure \ref{fig:rl_step}. We can find that at the first 1000 steps, all of the RL models can achieve a better results than the baseline. But rapid deterioration will occur for the GRPO and the Combined w/o sample filter when the training step is larger than 1500. This means that the DiffRO can be more stable with longer training steps.


\section{Conclusion}

This study explore the effect of RL for audio-based LLM including both ASR and TTS. 
We first propose an alternate GPU utilization framework which show great training efficiency based on audio input and output. 
For ASR, the study explores various rule-based reward functions using GRPO and investigates the impact of RL data construction. 
In TTS, it compares GRPO with DiffRO and demonstrates that combining both methods improves performance. Results show that RL significantly enhances both ASR and TTS systems, with reward design playing a crucial role in shaping model output preferences and user experience. 
This work highlights the potential of RL in audio-domain LLMs and calls for better-supported RL frameworks for audio applications.

\bibliographystyle{IEEEtran}
\bibliography{mybib}

\begin{thebibliography}{10}
\providecommand{\url}[1]{#1}
\csname url@samestyle\endcsname
\providecommand{\newblock}{\relax}
\providecommand{\bibinfo}[2]{#2}
\providecommand{\BIBentrySTDinterwordspacing}{\spaceskip=0pt\relax}
\providecommand{\BIBentryALTinterwordstretchfactor}{4}
\providecommand{\BIBentryALTinterwordspacing}{\spaceskip=\fontdimen2\font plus
\BIBentryALTinterwordstretchfactor\fontdimen3\font minus \fontdimen4\font\relax}
\providecommand{\BIBforeignlanguage}[2]{{%
\expandafter\ifx\csname l@#1\endcsname\relax
\typeout{** WARNING: IEEEtran.bst: No hyphenation pattern has been}%
\typeout{** loaded for the language `#1'. Using the pattern for}%
\typeout{** the default language instead.}%
\else
\language=\csname l@#1\endcsname
\fi
#2}}
\providecommand{\BIBdecl}{\relax}
\BIBdecl

\bibitem{ziegler2020finetuninglanguagemodelshuman}
\BIBentryALTinterwordspacing
D.~M. Ziegler, N.~Stiennon, J.~Wu, T.~B. Brown, A.~Radford, D.~Amodei, P.~Christiano, and G.~Irving, ``Fine-tuning language models from human preferences,'' 2020. [Online]. Available: \url{https://arxiv.org/abs/1909.08593}
\BIBentrySTDinterwordspacing

\bibitem{wu2025stepaudio2technicalreport}
\BIBentryALTinterwordspacing
B.~Wu, C.~Yan, C.~Hu, C.~Yi, C.~Feng, F.~Tian, F.~Shen, G.~Yu, H.~Zhang, J.~Li, M.~Chen, P.~Liu, W.~You, X.~T. Zhang, X.~Li, X.~Yang, Y.~Deng, Y.~Huang, Y.~Li, Y.~Zhang, Z.~You, B.~Li, C.~Wan, H.~Hu, J.~Zhen, S.~Chen, S.~Yuan, X.~Zhang, Y.~Jiang, Y.~Zhou, Y.~Yang, B.~Li, B.~Ma, C.~Song, D.~Pang, G.~Hu, H.~Sun, K.~An, N.~Wang, S.~Gao, W.~Ji, W.~Li, W.~Sun, X.~Wen, Y.~Ren, Y.~Ma, Y.~Lu, B.~Wang, B.~Li, C.~Miao, C.~Liu, C.~Xu, D.~Shi, D.~Hu, D.~Wu, E.~Liu, G.~Huang, G.~Yan, H.~Zhang, H.~Nie, H.~Jia, H.~Zhou, J.~Sun, J.~Wu, J.~Wu, J.~Yang, J.~Yang, J.~Lin, K.~Li, L.~Yang, L.~Shi, L.~Zhou, L.~Gu, M.~Li, M.~Li, M.~Li, N.~Wu, Q.~Han, Q.~Tan, S.~Pang, S.~Fan, S.~Liu, T.~Cao, W.~Lu, W.~He, W.~Xie, X.~Zhao, X.~Li, Y.~Yu, Y.~Yang, Y.~Liu, Y.~Lu, Y.~Wang, Y.~Ding, Y.~Liang, Y.~Lu, Y.~Luo, Y.~Yin, Y.~Zhan, Y.~Zhang, Z.~Yang, Z.~Zhang, B.~Jiao, D.~Jiang, H.-Y. Shum, J.~Chen, J.~Li, X.~Zhang, and Y.~Zhu, ``Step-audio 2 technical report,'' 2025. [Online]. Available: \url{https://arxiv.org/abs/2507.16632}
\BIBentrySTDinterwordspacing

\bibitem{schulman_proximal_2017}
\BIBentryALTinterwordspacing
J.~Schulman, F.~Wolski, P.~Dhariwal, A.~Radford, and O.~Klimov, ``\BIBforeignlanguage{en}{Proximal {Policy} {Optimization} {Algorithms}},'' Aug. 2017, arXiv:1707.06347 [cs]. [Online]. Available: \url{http://arxiv.org/abs/1707.06347}
\BIBentrySTDinterwordspacing

\bibitem{shao2024deepseekmathpushinglimitsmathematical}
\BIBentryALTinterwordspacing
Z.~Shao, P.~Wang, Q.~Zhu, R.~Xu, J.~Song, X.~Bi, H.~Zhang, M.~Zhang, Y.~K. Li, Y.~Wu, and D.~Guo, ``Deepseekmath: Pushing the limits of mathematical reasoning in open language models,'' 2024. [Online]. Available: \url{https://arxiv.org/abs/2402.03300}
\BIBentrySTDinterwordspacing

\bibitem{bai2024seedasrunderstandingdiversespeech}
\BIBentryALTinterwordspacing
Y.~Bai, J.~Chen, J.~Chen, W.~Chen, Z.~Chen, C.~Ding, L.~Dong, Q.~Dong, Y.~Du, K.~Gao, L.~Gao, Y.~Guo, M.~Han, T.~Han, W.~Hu, X.~Hu, Y.~Hu, D.~Hua, L.~Huang, M.~Huang, Y.~Huang, J.~Jin, F.~Kong, Z.~Lan, T.~Li, X.~Li, Z.~Li, Z.~Lin, R.~Liu, S.~Liu, L.~Lu, Y.~Lu, J.~Ma, S.~Ma, Y.~Pei, C.~Shen, T.~Tan, X.~Tian, M.~Tu, B.~Wang, H.~Wang, Y.~Wang, Y.~Wang, H.~Xia, R.~Xia, S.~Xie, H.~Xu, M.~Yang, B.~Zhang, J.~Zhang, W.~Zhang, Y.~Zhang, Y.~Zhang, Y.~Zheng, and M.~Zou, ``Seed-asr: Understanding diverse speech and contexts with llm-based speech recognition,'' 2024. [Online]. Available: \url{https://arxiv.org/abs/2407.04675}
\BIBentrySTDinterwordspacing

\bibitem{prabhavalkar2017minimumworderrorrate}
\BIBentryALTinterwordspacing
R.~Prabhavalkar, T.~N. Sainath, Y.~Wu, P.~Nguyen, Z.~Chen, C.-C. Chiu, and A.~Kannan, ``Minimum word error rate training for attention-based sequence-to-sequence models,'' 2017. [Online]. Available: \url{https://arxiv.org/abs/1712.01818}
\BIBentrySTDinterwordspacing

\bibitem{anastassiou_seed-tts_2024}
\BIBentryALTinterwordspacing
P.~Anastassiou, J.~Chen, J.~Chen, Y.~Chen, Z.~Chen, Z.~Chen, J.~Cong, L.~Deng, C.~Ding, L.~Gao, M.~Gong, P.~Huang, Q.~Huang, Z.~Huang, Y.~Huo, D.~Jia, C.~Li, F.~Li, H.~Li, J.~Li, X.~Li, X.~Li, L.~Liu, S.~Liu, S.~Liu, X.~Liu, Y.~Liu, Z.~Liu, L.~Lu, J.~Pan, X.~Wang, Y.~Wang, Y.~Wang, Z.~Wei, J.~Wu, C.~Yao, Y.~Yang, Y.~Yi, J.~Zhang, Q.~Zhang, S.~Zhang, W.~Zhang, Y.~Zhang, Z.~Zhao, D.~Zhong, and X.~Zhuang, ``Seed-{TTS}: {A} {Family} of {High}-{Quality} {Versatile} {Speech} {Generation} {Models},'' Jun. 2024, arXiv:2406.02430. [Online]. Available: \url{http://arxiv.org/abs/2406.02430}
\BIBentrySTDinterwordspacing

\bibitem{rafailov_direct_2023}
\BIBentryALTinterwordspacing
R.~Rafailov, A.~Sharma, E.~Mitchell, S.~Ermon, C.~D. Manning, and C.~Finn, ``\BIBforeignlanguage{en}{Direct {Preference} {Optimization}: {Your} {Language} {Model} is {Secretly} a {Reward} {Model}},'' May 2023. [Online]. Available: \url{https://arxiv.org/abs/2305.18290v3}
\BIBentrySTDinterwordspacing

\bibitem{du2025cosyvoice3inthewildspeech}
\BIBentryALTinterwordspacing
Z.~Du, C.~Gao, Y.~Wang, F.~Yu, T.~Zhao, H.~Wang, X.~Lv, H.~Wang, C.~Ni, X.~Shi, K.~An, G.~Yang, Y.~Li, Y.~Chen, Z.~Gao, Q.~Chen, Y.~Gu, M.~Chen, Y.~Chen, S.~Zhang, W.~Wang, and J.~Ye, ``Cosyvoice 3: Towards in-the-wild speech generation via scaling-up and post-training,'' 2025. [Online]. Available: \url{https://arxiv.org/abs/2505.17589}
\BIBentrySTDinterwordspacing

\bibitem{gao2025differentiablerewardoptimizationllm}
\BIBentryALTinterwordspacing
C.~Gao, Z.~Du, and S.~Zhang, ``Differentiable reward optimization for llm based tts system,'' 2025. [Online]. Available: \url{https://arxiv.org/abs/2507.05911}
\BIBentrySTDinterwordspacing

\bibitem{vonwerra2022trl}
L.~von Werra, Y.~Belkada, L.~Tunstall, E.~Beeching, T.~Thrush, N.~Lambert, S.~Huang, K.~Rasul, and Q.~Gallouédec, ``Trl: Transformer reinforcement learning,'' \url{https://github.com/huggingface/trl}, 2020.

\bibitem{sheng2024hybridflow}
G.~Sheng, C.~Zhang, Z.~Ye, X.~Wu, W.~Zhang, R.~Zhang, Y.~Peng, H.~Lin, and C.~Wu, ``Hybridflow: A flexible and efficient rlhf framework,'' \emph{arXiv preprint arXiv: 2409.19256}, 2024.

\bibitem{hu2024openrlhf}
J.~Hu, X.~Wu, Z.~Zhu, Xianyu, W.~Wang, D.~Zhang, and Y.~Cao, ``Openrlhf: An easy-to-use, scalable and high-performance rlhf framework,'' \emph{arXiv preprint arXiv:2405.11143}, 2024.

\bibitem{an2025funaudioasrtechnicalreport}
\BIBentryALTinterwordspacing
K.~An, Y.~Chen, C.~Deng, C.~Gao, Z.~Gao, B.~Gong, X.~Li, Y.~Li, X.~Lv, Y.~Ji, Y.~Jiang, B.~Ma, H.~Luo, C.~Ni, Z.~Pan, Y.~Peng, Z.~Peng, P.~Wang, H.~Wang, W.~Wang, W.~Wang, B.~Tian, Z.~Tan, N.~Yang, B.~Yuan, J.~Ye, J.~Yu, Q.~Zhang, K.~Zou, H.~Zhao, S.~Zhao, and J.~Zhou, ``Funaudio-asr technical report,'' 2025. [Online]. Available: \url{https://arxiv.org/abs/2509.12508}
\BIBentrySTDinterwordspacing

\bibitem{du2024cosyvoice2scalablestreaming}
\BIBentryALTinterwordspacing
Z.~Du, Y.~Wang, Q.~Chen, X.~Shi, X.~Lv, T.~Zhao, Z.~Gao, Y.~Yang, C.~Gao, H.~Wang, F.~Yu, H.~Liu, Z.~Sheng, Y.~Gu, C.~Deng, W.~Wang, S.~Zhang, Z.~Yan, and J.~Zhou, ``Cosyvoice 2: Scalable streaming speech synthesis with large language models,'' 2024. [Online]. Available: \url{https://arxiv.org/abs/2412.10117}
\BIBentrySTDinterwordspacing

\bibitem{r1aqa}
\BIBentryALTinterwordspacing
G.~Li, J.~Liu, H.~Dinkel, Y.~Niu, J.~Zhang, and J.~Luan, ``Reinforcement learning outperforms supervised fine-tuning: A case study on audio question answering,'' \emph{arXiv preprint arXiv:2503.11197}, 2025. [Online]. Available: \url{https://github.com/xiaomi-research/r1-aqa; https://huggingface.co/mispeech/r1-aqa}
\BIBentrySTDinterwordspacing

\end{thebibliography}
\end{document}